\documentclass[11pt,aps,prb]{revtex4}

\usepackage{graphicx}

\usepackage{epstopdf}	

\usepackage{lpic}
\usepackage{amsmath,amssymb}
\usepackage{nicefrac}
\usepackage[utf8]{inputenc}
\usepackage{hyperref}
\hypersetup{
    colorlinks=false,       
    colorlinks=true,
    linkcolor=blue,          
    citecolor=red,        
    filecolor=magenta,      
    urlcolor=cyan           
}
\usepackage{xcolor}
\usepackage{listingsutf8}
\newcommand{\ket}[1]{\left|{#1}\right\rangle}
\newcommand{\bra}[1]{\left\langle{#1}\right|}

\begin{document}
\title[Polarizing and controlling]{Robust techniques for polarization and detection of nuclear spin ensembles}

\author{Jochen Scheuer$^{1}$, Ilai Schwartz$^{2}$, Samuel Müller$^{1}$, Qiong Chen$^{2}$, Ish Dhand$^{2}$, Martin B. Plenio$^{2}$, Boris Naydenov$^{1}$\footnote{E-mail: boris.naydenov@uni-ulm.de} and Fedor Jelezko$^{1}$}
\affiliation{
$^1$Institute for Quantum Optics and Center for Integrated Quantum Science and Technology (IQST), Albert-Einstein-Allee 11, Universit\"at Ulm, 89069 Ulm, Germany\\
$^2$Institute of Theoretical Physics and Center for Integrated Quantum Science and Technology (IQST), Albert-Einstein-Allee 11, Universit\"at Ulm, 89069 Ulm, Germany
}


\begin{abstract}
Highly sensitive nuclear spin detection is crucial in many scientific areas including nuclear magnetic resonance spectroscopy (NMR), imaging (MRI) and quantum computing. The tiny thermal nuclear spin polarization represents a major obstacle towards this goal which may be overcome by Dynamic Nuclear Spin Polarization (DNP) methods. The latter often rely on the transfer of the thermally polarized electron spins to nearby nuclear spins, which is limited by the Boltzmann distribution of the former.  
Here we utilize microwave dressed states to transfer the high ($>$~92~\%) non-equilibrium electron spin polarization of a single nitrogen-vacancy center (NV) induced by short laser pulses to the surrounding $^{13}$C  carbon nuclear spins. The  NV is repeatedly repolarized optically, thus providing an effectively infinite polarization reservoir. A saturation of the polarization of the nearby nuclear spins is achieved, which is confirmed by the decay of the polarization transfer signal and shows an excellent agreement with theoretical simulations. Hereby we introduce the Polarization readout by Polarization Inversion (PROPI) method as a quantitative magnetization measure of the nuclear spin bath, which allows us to observe by ensemble averaging macroscopically hidden polarization dynamics like Landau-Zener-Stückelberg oscillations.
Moreover, we show that using the integrated solid effect both for single and double quantum transitions nuclear spin polarization can be achieved even when the static magnetic field is not aligned along the NV's crystal axis. 
This opens a path for the application of our DNP technique to spins in and outside of nanodiamonds, enabling their application as MRI tracers. Furthermore, the methods reported here can be applied to other solid state system where a central electron spin is coupled to a nuclear spin bath - e. g. phosphors donors in silicon and color centers in silicon carbide.
\end{abstract}

\maketitle

\section{Introduction}
Nuclear Magnetic Resonance (NMR) spectroscopy and Magnetic Resonance Imaging (MRI) are well established methods in chemistry, biology and medicine. Both techniques require the detection of the low magnetic field generated by thermally polarized nuclear spins, which demand high static magnetic fields and large sample quantities. Dynamical nuclear spin polarization (DNP) has been developed to transfer the much higher thermal electron spin polarization to the nuclear spins, thus improving the NMR signal. Another important application of DNP is in the emerging field of quantum information processing and especially for the initialisation of solid state quantum simulators \cite{Cai13, Suter15}. The improvement factor is usually limited by the ratio of the magnetic moments of the electron and nuclear spins, which determines the ratio of equilibrium polarization levels. Although recently \cite{Tateishi14} a much larger enhancement was demonstrated using optically polarized electron spins out of thermal equilibrium, the technique relies on short-living radical triplet states, which limits its applications. A very promising DNP method is based on optically pumped Nitrogen-Vacancy centers in diamond (NVs), where several demonstrations have been reported \cite{Jacques09, London13, Frydman13, Alvarez15, Pines15}, but they require very good alignment (below 1$^{\circ}$) of the magnetic field along the NV's crystal axis. Here we report and evaluate several methods for polarizing nuclear spins using a single NV center, that are applicable at arbitrary magnetic field strengths and wide range of orientations. Furthermore, we demonstrate a method to read out the magnetization of the nuclear spin bath - Polarization readout by Polarization Inversion (PROPI), where we obtain the number of spin quanta transferred from the NV center to the nuclear spins.

The NV is a unique physical system which became a universal sensor for magnetic \cite{Maze08,Gopi08} and electric fields \cite{Dolde11} and temperature \cite{Neumann13, Lukin13} with nanometer spatial resolution. Single NVs can be observed and the electron spin of the triplet ground state can be optically polarized and read out. Recently it has been demonstrated that the electron spin polarization can be transferred from an ensemble of NVs to the surrounding $^{13}$C nuclear spins in a bulk crystal using a specific experimental configuration, e. g. excited state level anti-crossing \cite{Frydman13}, ground state level anti-crossing or nearest neighbor $^{13}$C spins \cite{Alvarez15}. We have previously demonstrated by using proper microwave driving of the NV's electron spin, that polarization transfer can be achieved at arbitrary magnetic fields both for single \cite{London13} and ensemble of NVs \cite{Scheuer16}.

One of the main goals of diamond DNP is the polarization of nanodiamonds, which due to their bio-compatibility are well suited as hyperpolarized MRI tracers. However, as detailed in Ref.~\cite{Chen15}, the polarization of nanodiamonds via NV centers requires novel approaches for DNP, due to the random orientation of the NV center axes and its large zero-field splitting. A new theoretical DNP method has been proposed, combining the electron spin triplet $S = 1$ properties of the NV center, specifically its double-quantum transition (DQT), with the integrated solid effect (ISE) \cite{Henstra88a, Henstra14}. With these techniques an unprecedented level of robustness can be achieved, against misalignment of the externally applied static magnetic field with respect to the NV axis and against rotational diffusion of the NV.
Usually experimental validation of DNP protocols can only be achieved in a macroscopic ensemble of nuclear spins via NMR, thus limiting the information about the dynamics due to ensemble averaging. The accessibility of a single NV, including its control and detection, makes it a unique test bed for the effect of manipulation and polarization of the surrounding nuclear spin environment and for testing of novel DNP schemes. However, the readout of the nuclear spin state and polarization remained an open challenge, due to the limited quantifiable information in methods such as linewidth narrowing~\cite{London13}.
Here we demonstrate novel methods on a single NV center, which can be used not only to polarize, but also to measure the nuclear spin bath quantitatively.

In Section~\ref{PolarizationBuildUp} we present the NV's spin Hamiltonian and the general idea of our method. In Section~\ref{SQTExpImp} we present measurement results when using a single quantum transition. Experimental data for polarization via the NV's double quantum transition and for misaligned magnetic field are presented in Section~\ref{DQT}.

\section{Controlling the Nuclear Spin Bath}
\label{PolarizationBuildUp}
The physical system consisting of a single NV and a bath of $N^{\mathrm{^{13}C}}$ $^{13}$C nuclear spins can be described by the following Hamiltonian ($\hbar = 1$)
\begin{equation}
\hat H =  D \hat{S_z^2}+g\mu_B\vec{B}\cdot\vec{S}+\vec{S}\cdot\mathbf{A^{\mathrm{^{14}N}}}\cdot\vec{I}\,^{\mathrm{^{14}N}}+\sum_{j=1}^{j=N^{\mathrm{^{13}C}}}\gamma^{^{13}\mathrm{C}}\vert\vec B\vert\hat{I}_{jz}\,\!^{\mathrm{^{13}C}}+\vec{S}\cdot\sum_{j=1}^{j=N^{\mathrm{^{13}C}}}\mathbf{A_j^{\mathrm{^{13}C}}}\cdot\vec{I}_j\,\!^{\mathrm{^{13}C}}\mathrm{,}
\label{NVHamiltonian}
\end{equation}
where $D/2\pi=2.87\,$GHz is the zero field splitting of the ground state,  $g=2.003$ is the Land\'{e} factor, $\mu_B$ is the Bohr magneton, $\vec{B}=B_x\vec{e}_x+B_y\vec {e}_y+B_z\vec{e}_z$ is the applied static magnetic field, $\vec{S}=\hat S_x+\hat S_y+\hat S_z$ and $\vec{I}\,^{\mathrm{^{14}N}}=\widehat{I}_x\,\!^{\mathrm{^{14}N}}+\widehat {I}_y\,\!^{\mathrm{^{14}N}}+\widehat {I}_z\,\!^{\mathrm{^{14}N}}$ are the electron and nuclear spin operators of the NV ($S=1$ and for $^{14}$N $I=1$) and $\mathbf{A^{\mathrm{^{14}N}}}$ is the hyperfine interaction tensor of the NV. The Zeeman interaction of the $^{13}$C nuclear spin bath ($I=1/2$) is given by the second to last term in the Hamiltonian with their gyromagnetic ratio $\gamma^{^{13}\mathrm{C}}=6.728\times10^7$~T$^{-1}$s$^{-1}$. The last term describes the hyperfine interaction between the NV and the nuclear spin bath, where $\mathbf{A_j^{\mathrm{^{13}C}}}\sim \frac{g\mu_B\gamma^{^{13}\mathrm{C}}}{r_j^3}$ with $r_j^3$ the distance between the NV center and the nuclear spin. For simplicity we have removed the $^{14}$N nuclear spin Zeeman energy and its quadrupole interaction with the NV center. Unless otherwise stated in all experiments we set $\vert\vec{B}\vert\approx1750\,$G in order to compare the different measurements, though it should be noted that our methods can be applied at arbitrarily magnetic field strengths. The $z$ axis is taken to be along the NV crystal axis. MW pulses are applied on the electron spin resonance transition corresponding to NV nuclear spin state $\ket{m_I=+1}$. For the experiments presented in sections~\ref{SQTExpImp} and \ref{DQT} the magnetic field was aligned along $z$ better than $<0.5\,^\circ$.  Results for a misaligned field are presented in the last part of section~\ref{DQT}. We used various NV centers in different samples for our experiments, though for the results shown here the same NV center was used, in order to be able to compare quantitatively the different DNP methods.  The measurements were performed on a home built confocal microscope.

\begin{figure}[bht]
  \centering
  \includegraphics[width=1.0\textwidth]{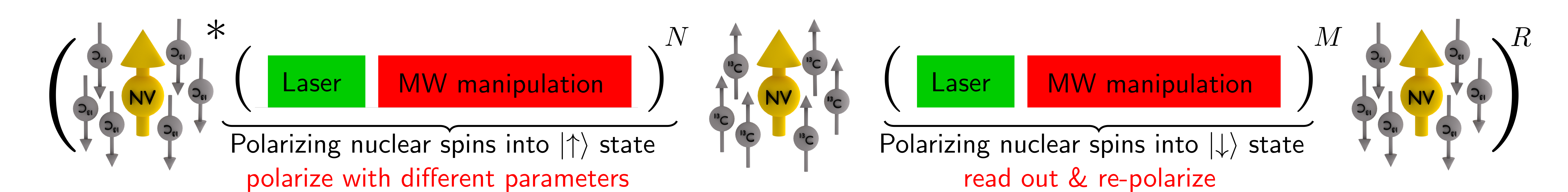}
\caption[Polarization sequence]{Schematic representation of the experiments. The first pulse sequence is repeated $N$-times and is used to polarize the nuclear spin bath along the static magnetic field ($\ket{\uparrow \uparrow \uparrow ...}$ state). By changing the MW manipulation, we can polarize the nuclear spin bath into the $\ket{\downarrow \downarrow \downarrow ...}$ state by repeating the sequence $M$-times. The illustrations present the state of the spin bath around the NV center at different time instants. At the beginning the nuclear spins are polarized in the $\ket{\downarrow}$ state. After repeating the polarization sequence $N$ times the nuclear spins are partially or fully inverted into the $\ket{\uparrow}$ state depending on $N$. By polarization of the nuclear spin bath during the $M$ cycles the nuclear spin bath is both initialized and read out. Hence, $M$ has to be set to larger enough, so that the nuclear spin bath is completely polarized. At the start of the experiment at $N,R=1$ (*) the nuclear spin bath is unpolarized, but due to larger number of repetitions of the measurements ($R = 5\cdot 10^4$) the first iteration can be neglected. For most of our experiments $N=50$ and $M=200$. See text for more details.}
\label{FigSequence}
\end{figure}

The schematic of the pulse sequence used in our experiments for polarizing the nuclear spin bath and reading out its magnetization is depicted in figure~\ref{FigSequence}.
It starts with a 3~$\mu$s long laser pulse to polarize the NV center into the $\ket{m_s=0}$ state, followed by a series of microwave (MW) pulses to transfer the polarization to the surrounding nuclear spins, which are coupled to the NV. The region where this hyperfine interaction is stronger than the coupling among the nuclear spins and is often referred as the ``frozen core" and also named ``spin diffusion barrier" \cite{Khuts62}. We will name it ``interaction region". Theoretical simulations show that in our experiments all nuclear spins with a coupling up to about 10 kHz to the NV center become completely polarized and nuclear spins coupled from 2 kHz to 10 kHz are partially polarized. Weakly coupled ones are not affected at all, but can be polarized via nuclear spin diffusion when the NV's electron spin is initialized into the $\ket{m_S=0}$ state.\\

The pulse sequence is repeated $N$ times, resulting in a polarization the nuclear spin bath in the $\ket{\uparrow \uparrow \uparrow ...}$ state, where the states $\ket{\uparrow}$ and $\ket{\downarrow}$ are defined to be respectively parallel and anti-parallel to the applied static magnetic field $\vec{B}$. In order to read out the nuclear spin bath, we apply another sequence, where we invert the nuclear spins into the $\ket{\downarrow}$ state, see figure~\ref{FigSequence}, right. By polarizing the nuclear spins in the opposite direction, compared to the previous sequence and reading out the NV state
optically, we are able to read out their magnetization quantitatively, see next section. In other words with the help of the NV center's electron spin we manipulate the nuclear spin environment and the back action of the latter on the NV is our signal. After normalizing the signal, we obtain the average number of spin quanta transferred from the NV to the nuclear spins, which is confirmed by theoretical simulations. We call this method Polarization readout by Polarization Inversion (PROPI). A typical experimental result after applying both pulse sequences is shown in figure~\ref{FigBuildUp}b \& c, see also next section.

\section{Polarization methods using single quantum transitions (SQT)}
\label{SQTExpImp}
\subsection{Nuclear Spin Orientation via Spin Locking (NOVEL)}
\begin{figure}[bht]
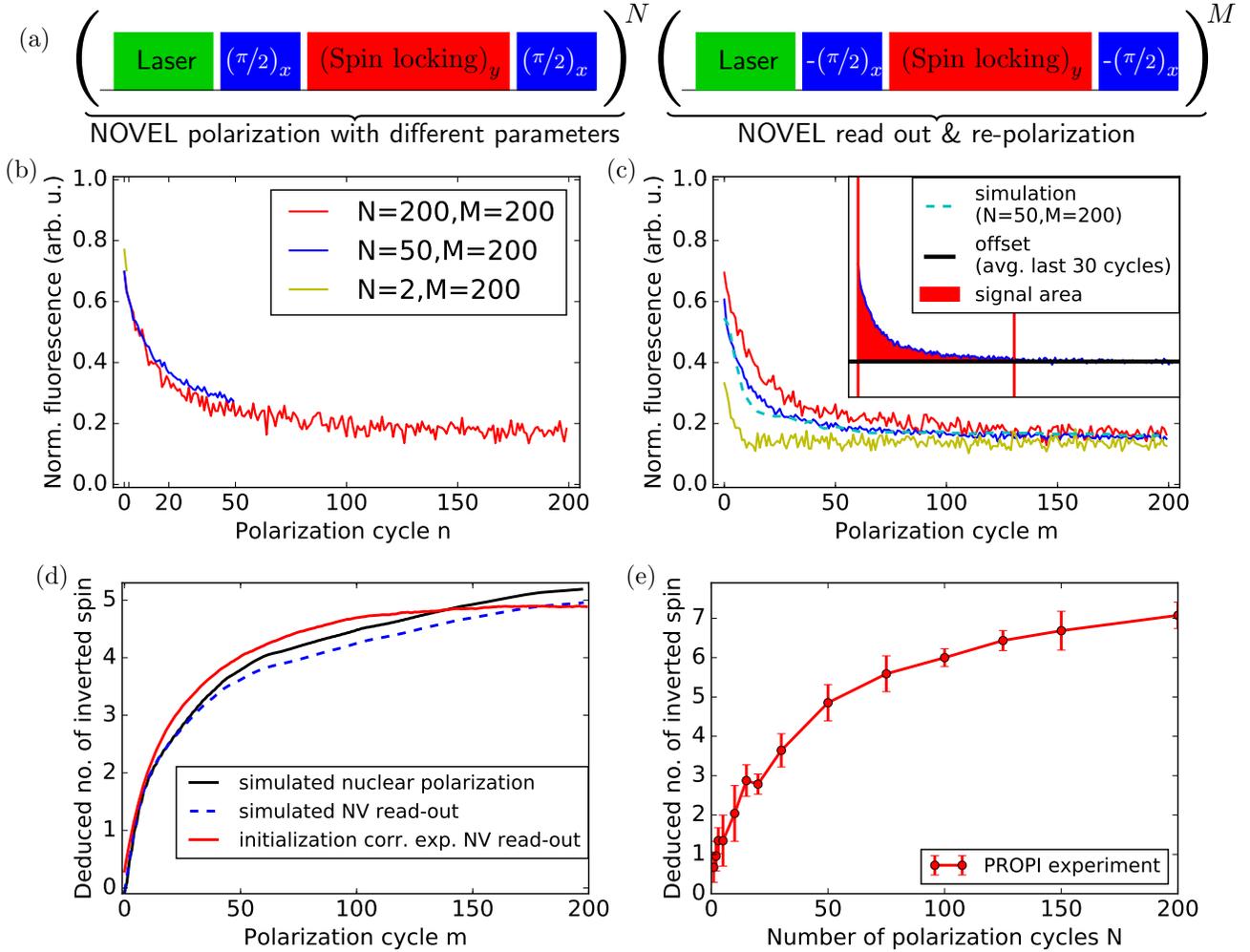

  \centering
\begin{lpic}{HaHa_Sequence(1.0,1.0)}
\lbl[t]{0,18;  (a)}
\end{lpic} 
 \begin{tabular}{cc}
 \begin{lpic}[l(-5mm)]{Polarisation_Dynamics_various_N(0.55,0.55)}
\lbl[t]{-3,100;  (b)} 
\end{lpic} &  
\begin{lpic}[l(-2.0mm)]{Polarisation_Dynamics_various_M(0.55,0.55)} 
\lbl[t]{-3,100;  (c)}
\end{lpic}  \\
\begin{lpic}[l(2.0mm)]{Polarisation_Dynamics_simulations(0.55,0.55)}
\lbl[t]{-3,100;  (d)} 
\end{lpic}&
\begin{lpic}[l(2.0mm)]{HaHa_buildup(0.55,0.55)}
\lbl[t]{-3,100;  (e)}  
\end{lpic} 
\end{tabular}  

\caption[Polarization Build Up]{{(a) Pulse sequences for the NOVEL experiment. (b) Fluorescence of the NV center measured with the $n$-th laser pulse, where $n=0,1,...N$ for $N=2$, (yellow), $N=50$ (blue) and $N=200$ (red), here $M=200$ for all data sets, see also figure~\ref{FigSequence}. (c) Fluorescence of the NV center measured with the $m$-th laser pulse, where $m=0,1,...M$. Inset: Measurement for $N=50$ and $M=200$. The red colored area enclosed by the two vertical lines is the PROPI signal, which gives the number of inverted nuclear spins and is a measure of the polarization. The dashed cyan curve is a simulation of the full experiment ($N=50; M=200$) (d) Comparison between simulated nuclear polarization, simulated NV read-out and initialization corrected PROPI data (see text) (e) Nuclear spin bath polarization as a function of the number of polarization cycles $N$, where $M=200$. }}
\label{FigBuildUp}
\end{figure}

First we demonstrate our method using the NOVEL DNP technique. Here the transfer of polarization from the NV's electron spin to the nuclear spins is realized via a spin locking experiment \cite{Henstra88, London13, Scheuer16}. The pulse sequence is depicted in figure~\ref{FigBuildUp}a. After initializing the NV in the $\ket{m_s=0}$ state, we align its spin along the $y$ axis of the Bloch sphere with a $\pi/2$ pulse, where it remains as long as the spin locking pulse is on (10~$\mu$s in our case) or until relaxation becomes important. During the spin locking pulse, the system is in the dressed state basis and the electron and nuclear spin flip-flop processes are energetically allowed. Afterwards we apply another $\pi/2$ pulse to bring the NV spin back to the $\ket{0}$, $\ket{1}$ basis, where its state is read out optically. The readout process is not part of the polarization protocol, but is used here to study the efficiency of the methods. A spin flip-flop process between the electron and nuclear spins occurs when the Hartmann-Hahn condition is fulfilled \cite{Hartmann62}
\begin{equation}
\Omega_1 = \omega^{^{13}\mathrm{C}}_0\mathrm{,}
\label{HartmannHahn}
\end{equation}
where $\Omega_1=g\mu_BB_1/\hbar$ is the Rabi frequency of the spin locking pulse with the magnetic field amplitude $B_1$ of the applied microwave. $\omega^{^{13}\mathrm{C}}_0=\gamma^{^{13}\mathrm{C}}B$ is the Larmor frequency of the $^{13}\mathrm{C}$ nuclear spins, see also equation~\ref{NVHamiltonian}.\\
In the experimental data shown in figure~\ref{FigBuildUp}b we observe that the NV's fluorescence level drops as a function of $N$ to a certain level, where it remains constant. The phase of the $\pi/2$ pulses around the spin locking time is the same, such that a bright fluorescence level indicates a spin flip during the spin locking time. The decay of fluorescence translates to a diminishing spin flip probability
(for $N\sim200$), meaning that the NV's polarization is completely transferred to the closest nuclear spins and no further electron-nuclear spin flip-flops can be observed. Hence the loss of fluorescence is a signature, that the nuclear spin bath is polarized. After that we invert this polarization by changing the phase of the $\pi/2$ pulses by 180 degrees and run the sequence for $M$ times. Now we again start to observe flip-flop processes, which are suppressed after $m\sim200$ cycles as expected (see figure~\ref{FigBuildUp}c), since the polarization transfer for both nuclear spin states must saturate for the same number of polarization cycles if we neglect spin diffusion to more distant nuclei. The area between the fluorescence signal and the offset (averaged over the last 30 points see Figure~\ref{FigBuildUp}c inset) indicates the number of spin quanta transfered from the NV to the nuclear spins. We obtain a quantitative measure of the nuclear spin polarization by correcting for imperfect NV initialization, arising from the charge state fluctuations leading to an initialization of 70\% NV$^-$\cite{Waldherr2011}, from the electron spin initialization of 92\% into the m$_s$=0 state\cite{Waldherr2011} and from initialization when the NV's nuclear spin is in a different state. This measure is the deduced number of inverted spins for the buildup of polarization with different values of $N$ as depicted in figure~\ref{FigBuildUp}e.
The accuracy of our measure is confirmed by theoretical simulations with  Hamiltonian \eqref{NVHamiltonian} having a nuclear spin bath of 30 spins (Figure~\ref{FigBuildUp} c, d) and using the time-evolving block decimation (TEBD) algorithm \cite{Prior10, Schollwoeck11,github_milan}. 
The nuclear spins are randomly sampled in a diamond lattice with natural abundance of $^{13}$C, the results are averaged over 30 different spin bath configurations and we account for above mentioned imperfect NV initialization. In figure ~\ref{FigBuildUp}d we compare the simulated polarization of the nuclear spins with the simulated NV fluorescence signal including all experimental parameters. These two together with the averaged experimental spin bath signal are in excellent agreement.
 
For $n=200$ and $m=200$ there is an offset (Figure~\ref{FigBuildUp}b,c) different from zero, which does not indicate polarization transfer. This effect can be explained by the driving of the off-resonant nitrogen hyperfine transitions. By defining the signal area between the fluorescence signal and the offset we solely observe the resonant, coherent nuclear-electron spin flip flop process. Due to the offset correction this directly translates to polarization transfer in terms of spin flip quanta (or quanta of angular momentum transferred). 
Effects like decoherence, pulse imperfections, spin relaxation and off-resonant driving will lower the efficiency of the polarization but they do not affect the accuracy of the readout. If this was the case, then they also influence the tail of the fluorescence signal and by subtracting the offset they are not observed in the signal (the area below first 100 pulses). If the polarization is not saturated during the tail (last 30 pulses) of the readout sequence, the offset level would be higher, meaning that the number of spin flip quanta would give a lower value than the actual transferred polarization. To avoid this in the following experiments the most effective polarization sequence (NOVEL) is chosen with a polarization cycle ratio $N$:$M$ of 50:200, such that the re-polarization part of PROPI ($M=200$ cycles) saturates the polarization to a high degree. The relaxation of the NV's electron spin in the ``dressed state basis" ($T_{1\rho}>200$ $\mu$s), thermal relaxation of the nuclear spin bath ($T_1$ process) and diffusion of the nuclear spin polarization out of the ``frozen core" are negligible at this time scale. In all experiments the  NV's fluorescence signal is normalized by a separate measurement of the electron spin population inversion via a $\pi$ pulse.\\

\begin{figure}[bht]
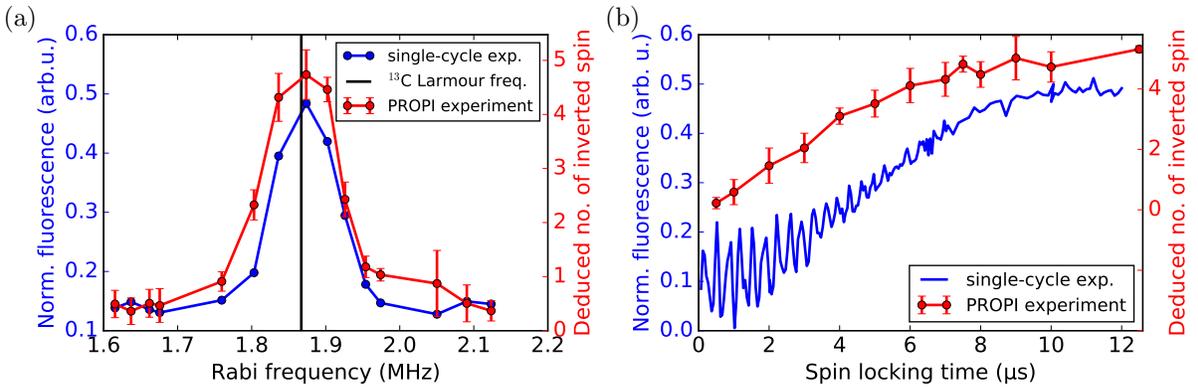

  \centering

 \begin{tabular}{cc}
 \begin{lpic}{HaHa_Drive_Pol_Res(0.5,0.5)}
\lbl[t]{-2,103;  (a)}
\end{lpic} &  
\begin{lpic}{HaHa_SL_Pol_Res(0.5,0.5)}
\lbl[t]{0,103;  (b)}
\end{lpic}
 \end{tabular} 
 
 \caption[Novel experiments]{(a) NOVEL experiment. Here the NV's fluorescence is observed as a function of the Rabi frequency of the spin locking pulse for a single polarization step where the nuclear spin bath is not polarized ($N=M=1$, blue curve). The red curve is a PROPI measurement which gives the average number of nuclear spin flips after polarizing the spin bath ($N=50$, $M=200$). An increase of the signal is observed when the condition~\ref{HartmannHahn} is fulfilled, which is a signature of polarization transfer. (b) NOVEL experiment where we measure how the signal depends on the length of the spin locking pulse for $N=M=1$ (blue curve) and $N=50$, $M=200$ (red curve). The oscillations in the single cycle measurement are due to coupling to the $^{14}$N nuclear spin of the NV.}

\label{FigNovel}

\end{figure}

In figure~\ref{FigNovel}a (blue curve) experimental data are shown, where the Rabi frequency $\Omega_1$ of the spin locking pulse is varied for a single polarization cycle and the NV's fluorescence is observed.

With this measurement we determine the correct Rabi frequency for controlling the nuclear spin bath. The PROPI experiment (red curve) uses $N=50 $ cycles where one spin locking parameter is varied and afterwards $M=200$ cycles with optimal spin locking parameters are applied. The latter 200 pulses are used to read out the magnetization and initialize the nuclear spin bath into the  $\ket{\downarrow}$ state. If we repeat the polarization sequence many times, the signal becomes broader (red curve). This is due to multiple spin flip-flop processes between the NV and different nuclear spins within the $N$ repetitions, resulting in a saturation  of polarization at a broader range of Rabi frequencies. In figure~\ref{FigNovel}b we plot the dependence of the NV's fluorescence on the length of the spin locking pulse, when equation~\ref{HartmannHahn} is fulfilled. The increase of the signal is again related to polarization transfer, which has a maximum at a length of the spin locking pulse of about 10 $\mu$s. Here again the blue curve represents a single polarization cycle ($N=M=1$), while the red curve was measured using $N=50$ and $M=200$. For the latter we observe a faster polarization transfer, which can be explained by the multiple spin flip-flops during the repetitions of the measurement. If we continue to increase the pulse length, the signal again decreases and shows oscillations (data not shown), as reported previously \cite{London13}. The signal of the single cycle experiment shows oscillations, which are not related to the $^{13}$C spins, but to hyperfine coupling to the $^{14}N$ nuclear spin. In all measurements the readout of the nuclear spin magnetization was performed by PROPI with the following parameters -  $M=200$, $\Omega_1 = \omega^{^{13}\mathrm{C}}_0$, spin locking time 10$\,\mu$s. In the further experiments we use the PROPI method as a benchmark to compare the efficiency of different polarization techniques.

\subsection{Integrated solid effect (ISE)}
\label{ISE}
We have previously demonstrated both experimentally \cite{Scheuer16} and theoretically \cite{Chen15} that by using the Integrated Solid Effect (ISE)\cite{Henstra88a} electron spin polarization from an ensemble of NVs can be transfered to the surrounding $^{13}$C nuclear spins. In this technique instead of a constant MW frequency, a frequency sweep over the electron spin resonance is used, which results in transfer of polarization. The effective Hamiltonian of this experiment considering an NV center coupled to a single nuclear spin can be written as \cite{Scheuer16}
\begin{equation}
H_{\mathrm{trans}} = \Omega\sigma_z+\Delta\sigma_x+B_{\mathrm{eff}}I_{z'}+\sigma_z (a_{z'} I_{z'}+a_{x'}I_{x'})
\label{ISEHamiltonian}
\end{equation}
with the effective Rabi frequency $\Omega=\Omega_M/\sqrt{2}$ and MW driving strength $\Omega_M=g\mu_BB_1/\hbar$, effective detuning $\Delta = D(\theta)-\gamma_eB-\delta(\theta)+\omega_M$ with $\theta$ the angle between the static magnetic field and the NV crystal axis and MW frequency $\omega_M$, second order correction $\delta(\theta)$, $\sigma_{x,z}$ and $I_{x',z'}$ the electron and nuclear spin operators in the rotated basis and $a_{z'}$ and $a_{x'}$ the secular and non-secular hyperfine interactions respectively.\\

\begin{figure}[bht]
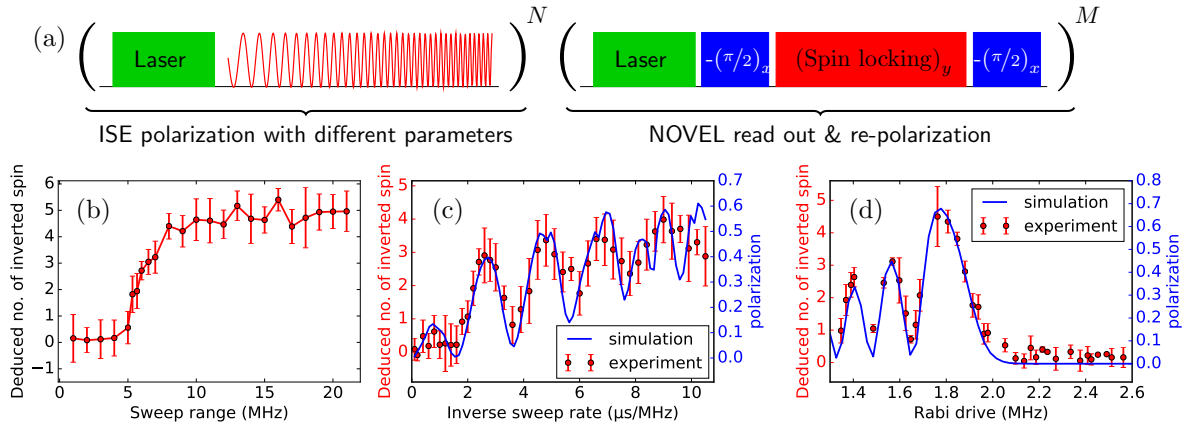
  
\centering
\begin{lpic}{sequence_ISE_HaHa(0.09,0.09)} 
\lbl[t]{0,190; (a)} 
\end{lpic} 
\begin{tabular}{ccc}
\begin{lpic}[l(4mm)]{ISE_range(0.34,0.34)} 
\lbl[t]{35,90; (b)}
\end{lpic} &  
\begin{lpic}[b(0.0mm),l(-2.0mm)]{ISE_speed_20times(0.34,0.34)} 
\lbl[t]{30,90; (c)}
\end{lpic}&  
\begin{lpic}[b(0.0mm),l(0.0mm)]{ISE_drive_20times(0.34,0.34)}
\lbl[t]{30,90; (d)}
\end{lpic} 
\end{tabular} 

\caption[SQT-ISE experiments]{DNP of the nuclear spin bath using ISE observed via PROPI. (a) Pulse sequence used in the experiments; (b) Dependence of the polarization on the sweep range of the MW. Transfer of polarization as a function of inverse sweep rate (c) and strength of the driving field (d). The red curves show the experimental data, the blue curves are simulations using three nuclear spins and without free parameters. The right axis present the sum of the $I_z$ of all three spins.}
\label{FigISE}
\end{figure}

In the ISE experiment there are three important parameters - the sweeping range of the MW frequency $f_{\mathrm{range}}$, the sweep speed $v=df/dt$ and the strength of the driving field $\Omega_M$. To find the optimal conditions is generally a complex problem, but some considerations help to narrow the range of spectral parameters. A detailed theoretical analysis of the method \citep{Chen15} reveals that for effective transfer of polarization the adiabatic condition with respect to the NV states has to be fulfilled while being only semi-adiabatic with respect to the flip-flop transition
\begin{equation}
\frac{\Omega^2}{\vert v\vert}\gg1 .
\end{equation}
On the other hand $v$ should not be too small to avoid adiabaticity for the flip-flop transition. In figure~\ref{FigISE} we show how the polarization transfer depends on these parameters.

In these experiments the nuclear spin bath was polarized using the ISE pulse sequence shown in figure~\ref{FigISE}a. It starts with the usual laser pulse to initialize the NV followed by a MW chirp pulse where its frequency is changed over the resonance transition. After that the polarization is measured and reinitialized by applying the NOVEL pulse sequence with optimal parameters for $M=200$. We observe that, by increasing the sweeping range, the polarization transfer improves until a saturation for $f_{\mathrm{range}}>10$~MHz is reached (figure~\ref{FigISE}b). The maximum of the nuclear spin polarization is the same as in the NOVEL experiment (figure~\ref{FigNovel}), showing that with both DNP methods similar nuclear spin polarization can be achieved. Due to the frequency sweep, ISE with the optimal parameters is about three times slower per polarization cycle than NOVEL, but it can be applied for broader spectral lines.
An increase of the spin polarization transfer is also observed when the inverse sweeping rate $1/\vert v\vert$ is increased (figure~\ref{FigISE}c) and additionally the signal oscillates. Similar effect is observed when the Rabi frequency is changed, see figure~\ref{FigISE}d, where the maximum polarization transfer is reached close to the Hartman-Hahn condition~(equation~\ref{HartmannHahn}).

The oscillating behavior in figure~\ref{FigISE},d is attributed to Landau-Zener-St{\"u}ckelberg oscillations \cite{Shevchenko10, Du11}, which is confirmed by the theoretical calculations (blue curves) using the Hamiltonian~\ref{ISEHamiltonian} with three nuclear spins, 20 polarization cycles (more cycles do not introduce changes) and no further free parameters. Such oscillations can only be observed in a single spin experiment, owing to the signal averaging in an ensemble experiment with many different nuclear spin environments.

\section{DNP using the Double Quantum Transition (DQT)}
\label{DQT}
\begin{figure}[bht]
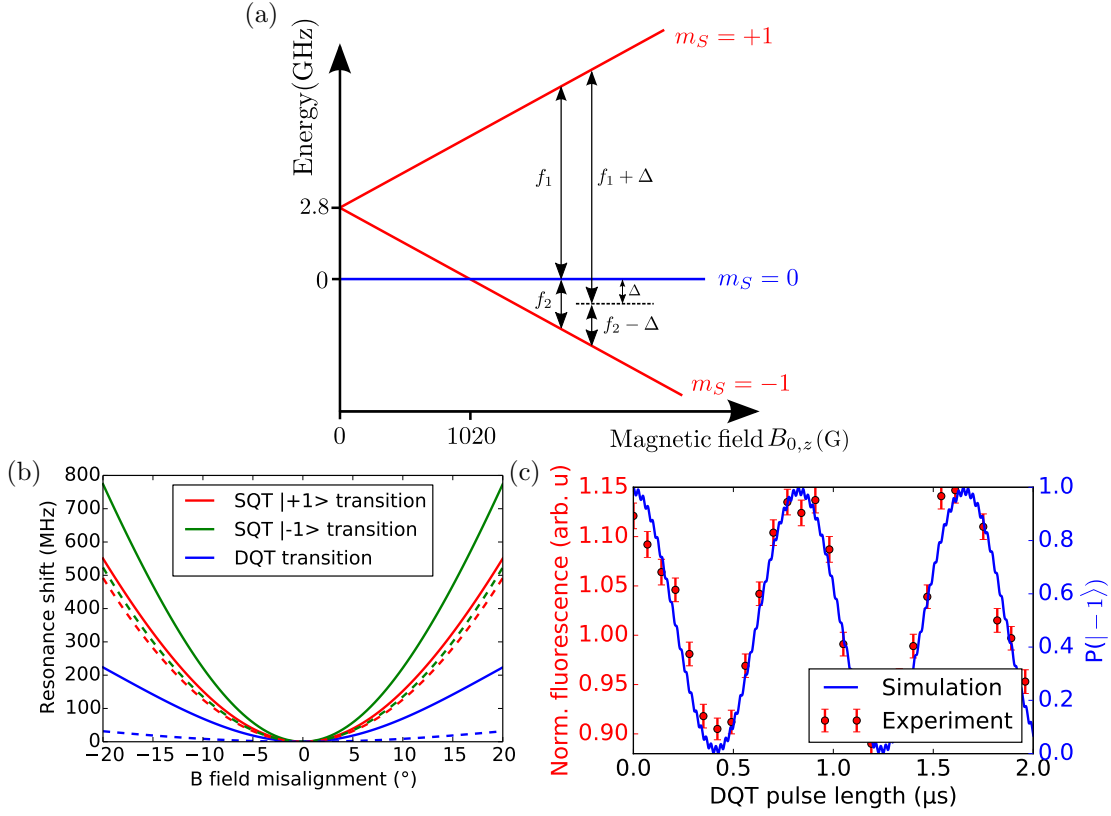

\centering
\begin{lpic}{Energy(0.4,0.4)}
\lbl[t]{-10,150; (a)}
\end{lpic}

\begin{tabular}{cc}
\begin{lpic}{lineshift1770GAnd1T(0.45,0.45)}
\lbl[t]{-2,100; (b)}
\end{lpic} &  
\begin{lpic}[b(-2.5mm)]{DQT_Rabi_simulation(0.45,0.45)}
\lbl[t]{-5,105.5; (c)}
\end{lpic}
\end{tabular}
 \caption[DQT Energy scheme]{ (a) Energy levels of the NV's electron as a function of the strength of the applied static magnetic field $\vec{B}$ along the NV's crystal axis. The arrows indicate the two MW frequencies $f_3=f1+\Delta$ and $f_4=f2-\Delta$ which are used to drive the double quantum transition $\ket{-1}\leftrightarrow\ket{+1}$. (b) Shift of the resonance line for the single and double quantum transitions as a function of the misalignment angle $\theta$ for 1770$\,$G (solid) and for 1$\,$T (dashed). (c) Experimental (blue markers) and theoretical (red curve, using the Hamiltonian~\ref{DQTHamiltonian}) DQT Rabi oscillations. }
 \label{FigDQTEnergies}
\end{figure}

One major drawback of NV centers in their application for DNP is that the electron spin polarization mechanism and its transition frequency depend on the orientation of the applied static magnetic field $\vec{B}$ with respect to the $z$ axis (angle $\theta$). In an ensemble of nanodiamonds each NV will have an arbitrary orientation, which may change over time and will lead both to significant broadening of the ESR line and to loss of ODMR contrast. In this case NOVEL cannot be used for DNP as it works in a narrow frequency range as shown in figure~\ref{FigNovel}a. Some of us have recently proposed \cite{Chen15} the use of the Double Quantum Transition ($\ket{-1}\leftrightarrow\ket{+1}$ shortly DQT) of NV centers in nanodiamonds at high magnetic fields (when $D\widehat{S}_z^2<<g\mu_B |\vec{B}|$) to reduce the sensitivity to misalignments. The DQT can be driven by applying simultaneously two MW frequencies which are detuned with $\Delta$ from the single quantum transitions, see figure~\ref{FigDQTEnergies}a.

Figure~\ref{FigDQTEnergies}b shows how the resonance frequency of SQT and DQT depends on the misalignment of $\vec{B}$. From this plot we can observe the stark difference in angle dependence between SQT and DQT. For example, for $B=1\,$T and $\theta = 20^\circ$ the resonance shift in DQT is less than 50 MHz, compared with SQT where it is about 500 MHz. A typical experimental data set of a DQT Rabi measurement is depicted in figure~\ref{FigDQTEnergies}c, where the signal can be simulated using the Hamiltonian~\ref{DQTHamiltonian} explained below.

Next we describe the DQT DNP method. Under MW irradiation with frequencies $\omega^{MW}_{-1}=2\pi f_1$, $\omega^{MW}_{1}=2\pi f_2$ and amplitudes (Rabi frequencies) $\sqrt{2}\,\Omega_1, \sqrt{2}\,\Omega_{-1}$, the following Hamiltonian in the interaction picture is obtained
\begin{equation}
\label{DQTHamiltonian}
H = \frac{\delta}{2} S_z + \Delta S_z^2 + \frac{\Omega_1}{2}(\ket{1}\bra{0} + \ket{0}\bra{1}) + \frac{\Omega_{-1}}{2}(\ket{-1}\bra{0} + \ket{0}\bra{-1}) +\gamma_I B I_{z'}+S_z (a_{z'} I_{z'}+a_{x'}I_{x'})
\end{equation}
Defining $\Delta_1 = \omega_1 - \omega^{MW}_1$, $\Delta_{-1} = \omega_{-1} - \omega^{MW}_{-1}$, then $\delta = \Delta_1+\Delta_{-1}$ is the detuning from the DQT, and $\Delta = (\Delta_1 - \Delta_{-1}) / 2$ is the detuning from the $\ket{0}$ state, see figure~\ref{FigDQTEnergies}a.
As $\Delta > \Omega_1, \Omega_{-1}$, the $\ket{0}$ state is detuned from the dynamics and we obtain an effective Hamiltonian for the $\{\ket{-1},\ket{1}\}$ subspace after adiabatic elimination of the $\ket{0}$ state
\begin{equation}
\label{DQTEffHamiltonian}
H_\mathrm{DQT} = (\delta+\delta_{so}) \sigma_z + \Omega_\mathrm{eff} \sigma_x + \gamma_I B I_z + 2\sigma_z (a_{z'} I_{z'}+a_{x'}I_{x'})
\end{equation}
with $\sigma_z = \frac{1}{2}(\ket{1}\bra{1} - \ket{-1}\bra{-1})$, $\sigma_x = \frac{1}{2}(\ket{1}\bra{-1} + \ket{-1}\bra{1})$ and $\delta_{so}= \frac{ \Omega_1^2 - \Omega_{-1}^2}{4\Delta}$ is the detuning due to second order corrections for $\Delta \gg \Omega_1, \Omega_{-1}$. The DQT Rabi frequency when $\Omega_1 \Omega_{-1}=\Omega_{\mathrm{SQT}}$ and $\delta=0$ is given by $\Omega_{eff}= \nicefrac{1}{2} (\sqrt{2\alpha \Omega^2_{\mathrm{SQT}} +\Delta^2}-\Delta)$, where typical experimental parameters are $\Omega_{\mathrm{SQT}} \approx 10\,$MHz and $\Delta = 40\,$MHz. The factor $\alpha$ is introduced for convenient sweeping of the Rabi frequency, see figure~\ref{FigDQTISE}d.

Time evolution under the effective Hamiltonian~\ref{DQTEffHamiltonian} will result in polarization transfer from the NV center to a single nuclear spin. Here the effective hyperfine coupling is two times larger compared to the SQT experiments, see equation~\ref{ISEHamiltonian}. This result is confirmed by the experiment as shown below.

\subsection{NOVEL using DQT}
\begin{figure}[htb]
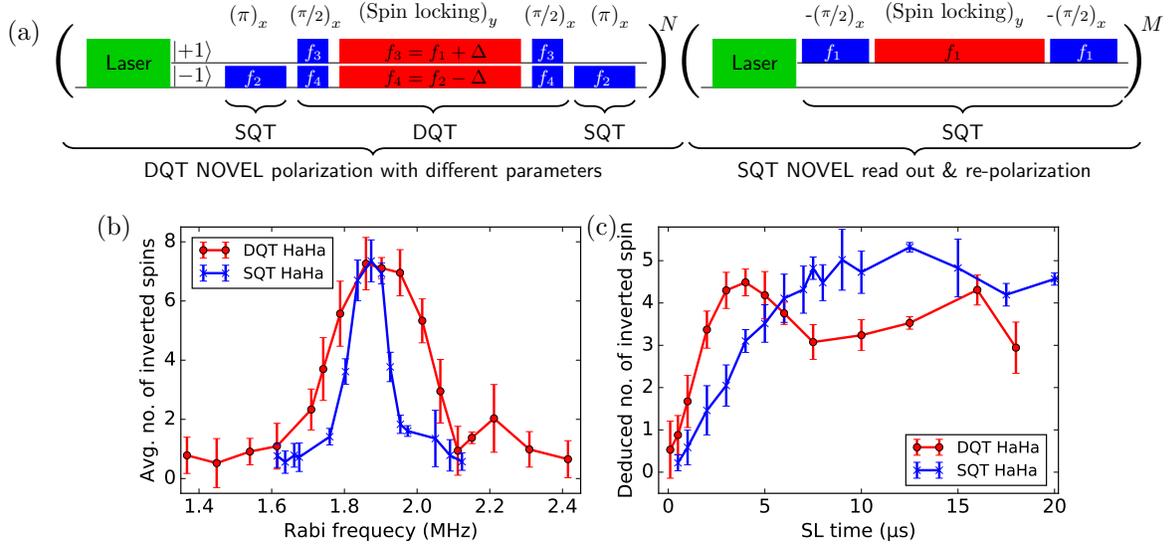

\centering
\begin{lpic}{DQTHaHa_Sequence(0.8,0.8)}
\lbl[t]{0,28; (a)}
\end{lpic}

\vspace{3mm}

\begin{tabular}{cc} 
\begin{lpic}{DQT_HaHa_Driving(0.45,0.45)}
\lbl[t]{-5,99; (b)} 
\end{lpic} &   
\begin{lpic}{DQT_HaHa_SL(0.45,0.45)}
\lbl[t]{-2,99; (c)}
\end{lpic} 
\end{tabular} 

\caption[DQT Haha]{(a) Pulse sequence for the DQT-NOVEL experiment. Efficiency of the transfer of polarization for SQT (blue) and DQT (red) as a function of the Rabi frequency (b) and the length of the spin locking pulse (c). All experiments are measured using PROPI with optimal NOVEL conditions.}
\label{FigDQTNovel}
\end{figure}

The pulse sequence for the NOVEL experiment using the DQT is shown in figure~\ref{FigDQTNovel}a.
After the initialization of the NV in $\ket{0}$, we apply a $\pi$ pulse on the $\ket{0}\leftrightarrow\ket{-1}$ transition, followed by two pulses on both transitions, which serve as a DQ $\pi/2$ pulse. Afterwards we apply a DQ spin locking pulse, during which the NV's electron spin is in the dressed state basis. Then we transform back to the $\ket{0},\ket{-1},\ket{1}$ basis. The magnetization of the nuclear spin bath is read out by PROPI using a SQT-NOVEL sequence with $M=200$. In figure~\ref{FigDQTNovel}b we compare the efficiency of the polarization transfer for SQT-NOVEL (blue curves) and DQT-NOVEL (red curves) as a function of the Rabi frequency. We observe, that for the DQT measurement spin flip-flop processes occur for a wider range of MW powers, compared to SQT-NOVEL. By increasing the length of the spin locking pulses, DQT reaches maximum earlier compared to SQT as shown in figure~\ref{FigDQTNovel}b. Both signals show oscillations, which are caused by the $^{13}$C spins with the strongest coupling (compared to the rest of the nuclear spin bath) to the NV. The DQT shows the doubled frequency (60~kHz) compared to SQT (30~kHz), as expected from the theory (see equation~\ref{DQTEffHamiltonian}) due to the twice larger magnetic dipole moment.

\subsection{ISE using DQT}
\begin{figure}[htb]
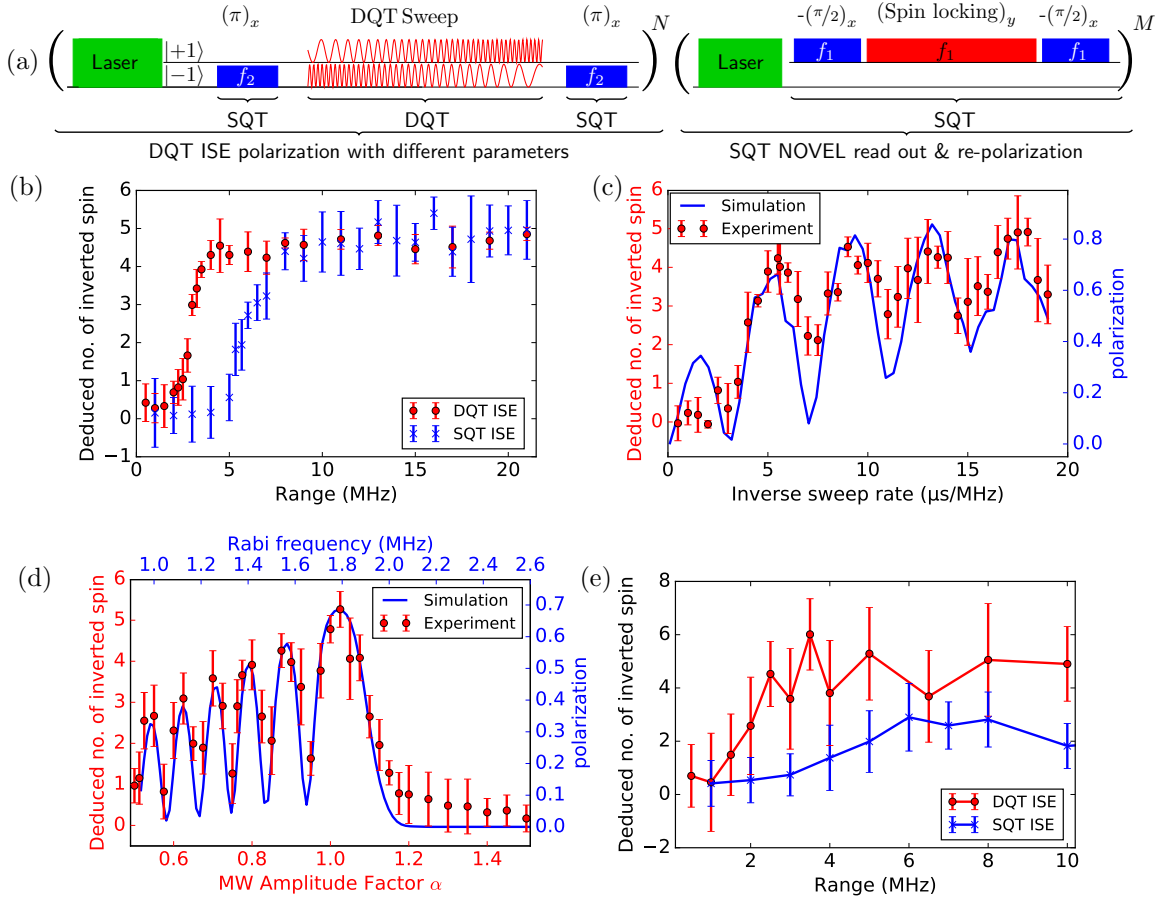

\centering 
\begin{lpic}{DQTISE_Sequence(0.08,0.08)}
\lbl[t]{0,200; (a)}
\end{lpic}
\begin{tabular}{cc}
\begin{lpic}[l(-5.0mm)]{DQT_ISE_Range(0.45,0.45)}
\lbl[t]{-13,100; (b)}
\end{lpic}&
\begin{lpic}[b(-0.0mm)]{DQT_ISE_SweepRate(0.45,0.45)}
\lbl[t]{-3,100; (c)}
\end{lpic}\\
\begin{lpic}[l(3.5mm)]{DQT_ISE_Amplitude(0.45,0.45)}
\lbl[t]{-13.5,100; (d)}
\end{lpic}&
\begin{lpic}[b(0.0mm),l(-7.5mm)]{DQTISE_ISE_misaligned(0.45,0.45)}
\lbl[t]{-4,100; (e)}
\end{lpic} 
\end{tabular} 
\caption[DQT ISE]{(a) Pulse sequence for the DQT-ISE experiment. (b) Polarization transfer as a function of the frequency range for SQT-ISE (blue) and DQT-ISE (red). Polarization transfer dependence on the sweeping rate (c) and MW amplitude factor $\alpha$ proportional to the Rabi frequency (d) for DQT (red markers). The blue curves are simulations. (e) Misalignment of the external magnetic field in respect to the NV axis of 5$^{\circ}$; performance of DQT-ISE (red) and SQT-ISE (blue) depending on the range of the sweep (here: $N=50$ and $M=300$). Readout is performed by PROPI with Hartmann-Hahn conditions matched.} 
\label{FigDQTISE}
\end{figure}

In the previous section we have shown polarization build up via the DQT-NOVEL method, but the resonance frequency range is limited (see figure~\ref{FigDQTNovel}b). Here we implement ISE  using the DQT, where both larger frequency range and robustness against misalignment of the magnetic field is expected. The pulse sequence used in the experiment is depicted in figure~\ref{FigDQTISE}a.

Again the NV is put into the $\ket{-1}$ state by a laser and a $\pi$ pulse. Simultaneously two chirp pulses are applied on both NV transitions. The frequency of the pulse on the $\ket{0}\leftrightarrow\ket{+1}$ transition is increasing, while on the other one it is decreasing. Here $\Delta$ is kept constant, while $\delta$ is changed, see equation~\ref{DQTHamiltonian}. At the end we use again a $\pi$ pulse on the $\ket{0}\leftrightarrow\ket{-1}$ transition and then read out the NV state. The nuclear spin bath magnetization is read out by PROPI using a SQT-NOVEL sequence. First we compare how the polarization transfer depends on the range of the frequency sweep for DQT and SQT, figure~\ref{FigDQTISE}b. We find that for the former measurement a saturation is reached for about half of the range compared to the SQT, which could be explained by the doubled effective hyperfine interaction constant to the $^{13}$C nuclear spins. When we change the sweeping rate (see figure~\ref{FigDQTISE}c), we again observe the Landau-Zehner-St{\"u}ckelberg oscillations, similar to the SQT-ISE experiment (see also figure~\ref{FigISE}b). Analogous to SQT-ISE is also the result of changing the DQ Rabi frequency shown in figure~\ref{FigDQTISE}d. This behavior can be reproduced with simulations (blue curves in figures~\ref{FigDQTISE}c,d) using the Hamiltonian~\ref{DQTEffHamiltonian} and same parameters as in section~\ref{ISE} adding additional MW power fluctuations of $\pm 5\%$.

At high magnetic fields when the magnetic field is not aligned with the NV crystal axis, both the optical polarization and contrast of the readout of the NV center are significantly reduced \cite{Jacques09}. However, our readout method still enables the measurement of the nuclear bath polarization though  with low contrast. 

The misalignment shifts the NV's resonance frequency by $\Delta_{\mathrm{SQT}}$ leading to the effective Rabi frequency $\Omega_{eff}= \sqrt{ \Omega^2_{\mathrm{SQT}} + \Delta^2_{\mathrm{SQT}}}$, where the MW drive $\Omega_{\mathrm{SQT}}$ is constant. From figure \ref{FigNovel} we observe that the width of the Hartmann-Hahn resonance is approximately 100~kHz in the SQT. At the applied magnetic field of 1770~G and a misalignment angle of 5$^{\circ}$ with the NV crystal axis the electron spin resonance is shifted by $\Delta_{\mathrm{SQT}} \approx 60\,$MHz, resulting in $\Omega_{eff}\approx60\,$~MHz since $\Omega_{\mathrm{SQT}}=1.86\,$~MHz. Here we are far away from the Hartmann-Hahn condition~\ref{HartmannHahn}.

We have performed DNP experiments with $\theta=5^{\circ}$. In figure~\ref{FigDQTISE}e we compare SQT ISE and DQT ISE showing how the polarization transfer depends on the frequency sweep range, which is centered around the new, misaligned resonance condition. We observe similar behavior compared to the aligned case. Beginning from a certain range, polarization is transfered from the NV to the nuclear spins, where for DQT ISE the required range is shorter than with SQT ISE.
In order to measure this polarization transfer PROPI is used  with optimal parameters of the NOVEL sequence. The signal obtained in the experiments is rather low, compared to the signal when the magnetic field is aligned. This effect is not due to low efficiency of the polarization transfer, but rather a result of the inefficient readout of the nuclear spin bath. Owing to misalignment induced loss of optical polarization efficiency and no nitrogen hyperfine spin state polarization\cite{neumann2012towards} the ODMR contrast is approx. 10 fold reduced. These lead also to lower polarization efficiency in NOVEL which means that compared to the aligned case more polarization cycles ($M=300$) are needed to reinitialize the spin bath. Due to the low signal the measurement time increased 100 times, thus allowing us to perform an experiment only for one off-axis angle of the magnetic field.

In addition to requiring a shorter sweep range as in the aligned case, DQT-ISE enables covering a larger angle misalignment for the same sweep range. As shown in figure~\ref{FigDQTEnergies}c, in 1~T magnetic field, a 5 degree angle will correspond to barely a few MHz shift in the DQT resonance, compared to over 30 MHz for the SQT. 

\section*{Conclusions}
We presented a novel method for reading out bath's magnetization named Polarization readout by Polarization Inversion (PROPI) providing a measure of polarization of the nearby nuclear spins. With this technique we have evaluated new robust methods for controlling the nuclear spin bath ($^{13}$C nuclear spins) surrounding a central electron spin (single NV center). In contrast to previous reports, our polarization techniques can be applied at a wide range of magnetic fields and sample orientations. This enables the implementation of optical DNP in nanodiamonds and also in $^{13}$C enriched samples with broad spectral line widths.  We compare the performance of the different methods, where the experimental data are well supported by theoretical simulations. The methods demonstrated here could be applied to other promising systems like color centers in silicon carbide, where nuclear spin polarization has been also observed \cite{Falk15}. We believe that the results reported here will find an application in DNP NMR spectroscopy and MRI with chemically modified nanodiamonds as markers.

\section*{Acknowledgements}

We thank Liam P. McGuinness, Thomas Unden, Simon Schmitt, Paz London, Xi Kong, Benedikt Tratzmiller and Christoph M\"uller for fruitful discussions and technical support. We thank Jan Haase for providing code to obtain hyperfine coupling constants in the diamond lattice.
This work was supported by the ERC Synergy grant BioQ, the EU (DIADEMS, HYPERDIAMOND grant agreement No 667192), the DFG (SFB TR/21, FOR 1493) and the Volkswagenstiftung. BN is grateful to the Postdoc Network program of the IQST and to the Bundesministerium f\"ur Bildung und Forschung for receiving the ARCHES award. IS acknowledges a PhD fellowship from the IQST. I.D. acknowledges financial support from the Alexander von Humboldt Foundation via the Humboldt Research Fellowship for Postdoctoral Researchers.



\section*{References} 
 
%
\end{document}